\documentclass[runningheads]{llncs}

\usepackage[T1]{fontenc}
\usepackage{graphicx}
\usepackage{booktabs}
\usepackage{amsmath}
\usepackage{xcolor}
\usepackage{url}
\usepackage{hyperref}
\usepackage{multirow}
\usepackage{enumitem}
\usepackage{amssymb}
\usepackage{tikz}
\usepackage{pgfplots}
\usepackage{kotex}
\usepackage{xstring}
\usepackage{microtype}
\usepackage[none]{hyphenat}
\usepackage{makecell}
\pgfplotsset{compat=1.18}
\usetikzlibrary{positioning, arrows.meta, calc, decorations.pathreplacing, patterns}

\newcommand{\PP}[1]{
\vspace{2px}
\noindent{\bf \IfEndWith{#1}{.}{#1}{#1.}}
}

\begin{document}

\title{Poisoned Playbooks: Demystifying Knowledge Poisoning Effects on AI Security Agents}
\titlerunning{Poisoned Playbooks}

\author{Juho Park\inst{1}\orcidID{0009-0003-4394-0449} \and
Hyunmin Choi\inst{2,*}\orcidID{0009-0002-0486-9582} \and
Kevin Nam\inst{3,*}\orcidID{0000-0002-4621-2434}
}

\authorrunning{J. Park, H. Choi, and K. Nam}
\institute{Security Defense, NAVER Cloud, Gyunggi-do, Republic of Korea \and
College of AI Convergence, Dankook University, Gyunggi-do, Republic of Korea \and
School of Computing, Kyung Hee University, Gyunggi-do, Republic of Korea
\email{juho.park@navercorp.com, hyunmin.choi@dankook.ac.kr, kvnam@khu.ac.kr}\\
*: correspondence should be addressed to H. Choi and K. Nam
}

\maketitle

\begin{abstract}
AI security agents increasingly rely on Retrieval-Augmented Generation (RAG) to use external security knowledge for vulnerability analysis and exploit reasoning. This creates a new risk: poisoned write-ups can be operationalized into incorrect exploit behavior.
Yet, prior work on RAG poisoning has mostly studied answer corruption in QA settings, much less is known about action-taking security agents.
This paper aims to reveal such characteristics with crafted poisons about real-world challenges and AI agents.
First, we demonstrate how a crafted single poisoned write-up injected into public-style security knowledge sources which we denote as \textit{Poisoned Playbooks}, alters the behavior of RAG-based AI security agents.
Across 11 CTF challenges, 3 frontier LLM families, 2 model generations, and 11 real-world CVEs, we find that poison adoption is systematic rather than random.
To explain this pattern, we introduce the \emph{Verification Boundary} (VB), a 3-level empirical classification based on what evidence the agent can use to refute a retrieved claim.
Finally, we evaluate verification prompting and multi-source retrieval, showing that both help when stronger evidence exists, but weaken under sparse-evidence and zero-day conditions.

\end{abstract}
\section{Introduction}\label{sec:intro}

Large language models (LLMs)~\cite{touvron2023llama,achiam2023gpt} are increasingly deployed not only as conversational assistants, but also as agents that retrieve external information, reason over it, and act on behalf of users. A common architectural pattern in such systems is Retrieval-Augmented Generation (RAG)~\cite{lewis2020rag,gao2024rag,karpukhin2020dense,huang2024survey}, which supplements parametric model knowledge with external documents retrieved at inference time. While RAG improves freshness and domain specificity, it also creates a new integrity risk--recent works demonstrated that malicious documents can corrupt RAG pipelines using numerous methods~\cite{poisonedrag2025,confundo2025,phantom2024}.

This paper focuses on how such attacks can affect AI security agents.
Security is a particularly important example of such a setting. LLM-based systems are increasingly being explored for incident investigation, threat hunting, vulnerability analysis, exploit reasoning, and autonomous penetration testing~\cite{fang2024llm,xbow2024,liu2026kryptopilot,happe2023getting}. In these workflows, external knowledge is consumed as operational guidance. A retrieved write-up may shape not only an explanation, but also a concrete exploit hypothesis, a sequence of commands, or a decision to abandon one attack path in favor of another. As a result, poisoning in this domain threatens the behavior of the agent itself, not merely the text it produces.
The risk is amplified by the way \textit{Security Knowledge} (SK) is produced and consumed.
In practice, analysts often rely on fast-moving public sources such as exploit repositories, community vulnerability templates, GitHub repositories, issue threads, proof-of-concept releases, and blog-based write-ups to understand newly disclosed vulnerabilities and their exploitation conditions~\cite{ji2025ctf,exploitdb,nuclei}. While these sources are timely and useful, they are also noisy, weakly curated, and susceptible to manipulation; prior work has shown that public exploit repositories and GitHub-hosted PoCs can contain misleading or even malicious content~\cite{elyadmani2025}. This problem is especially acute in sparse-evidence settings such as newly emerging or zero-day vulnerabilities, where authoritative counter-evidence may be unavailable or delayed.

Motivated by this gap, we study how poisoned external knowledge affects the behavior of AI security agents. Our vehicle is \emph{Poisoned Playbooks}: crafted write-ups injected into public-style SK repositories and then consumed by RAG-based agents during vulnerability-solving tasks. Unlike prior poisoning settings centered on factual QA~\cite{poisonedrag2025,confundo2025}, our focus is action-oriented: the agent must convert retrieved claims into exploit strategies and concrete decisions.
To structure this study, we investigate the following research questions:

\begin{itemize}[wide, nosep, labelwidth=!, labelindent=0pt]
    \item[\textbf{RQ1.}] \textit{Can a single poisoned write-up injected into a public SK source systematically alter the behavior of RAG-based AI security agents?} (\S\ref{sec:outcomes})
    \item[\textbf{RQ2.}] \textit{What determines whether a poisoned claim is rejected or adopted, and does this pattern generalize across model families, model generations, and real-world vulnerabilities?} (\S\ref{sec:outcomes}, \S\ref{sec:vb}, \S\ref{sec:cve})
    \item[\textbf{RQ3.}] \textit{Which mitigation strategies reduce poison adoption, and under what conditions do they fail?} (\S\ref{sec:defense})
\end{itemize}

We answer these questions through a study spanning 11 CTF challenges, three frontier LLM families, two model generations, and 11 real-world CVEs. Our main finding is that poison adoption is governed less by vulnerability difficulty alone than by what evidence the agent can use to independently challenge a retrieved claim. We capture this pattern through the \emph{Verification Boundary} (VB), a three-level empirical classification that separates claims that can be checked against local artifacts, claims that depend on model-side knowledge, and claims whose refutation lies outside the agent's immediate observable context.
In brief, our contributions are as follows:
\begin{itemize}[wide, nosep, labelwidth=!, labelindent=0pt, label=$\bullet$]
  \item We present a systematic empirical study of how knowledge poisoning affects AI security agents, spanning 11 CTF challenges, 3 frontier LLM families, 2 model generations, and 11 real-world CVEs.
  \item We present the \emph{Verification Boundary}, a 3-level classification that explains when retrieved claims are rejected, model-dependent, or consistently adopted.
  \item We show that the same pattern generalizes from controlled challenge settings to real-world CVEs, and that sparse-evidence regimes remain especially vulnerable.
  \item We evaluate practical mitigation strategies and identify the conditions under which they help and the regimes in which they remain insufficient.
\end{itemize}
\section{Background}\label{sec:background}

\subsection{Retrieval-Augmented Generation}

Retrieval-Augmented Generation (RAG) augments a large language model (LLM) with an external knowledge source, allowing the model to condition its outputs on retrieved documents rather than relying solely on parametric memory~\cite{lewis2020rag,gao2024rag}. A standard RAG pipeline consists of a knowledge database, a retriever, and an LLM: the retriever selects relevant documents for a query, and the LLM generates its response conditioned on the retrieved context~\cite{poisonedrag2025}. This architecture improves freshness and domain specificity without requiring model retraining.

At the same time, RAG expands the system's trust boundary. In a standalone LLM, integrity concerns primarily involve the model parameters and prompt context. In a RAG system, correctness also depends on the integrity of the external corpus and the retrieval process that selects from it. The knowledge database therefore becomes part of the system's effective decision surface~\cite{poisonedrag2025}.

\subsection{AI-Powered Security Agents and SK Sources}

Recent advances in LLM reasoning and tool use have enabled AI-powered security agents that combine language-model inference with tools such as shell access, web interaction, and code execution. These systems are increasingly explored for incident investigation, threat hunting, vulnerability analysis, exploit reasoning, and autonomous penetration testing~\cite{fang2024llm,enigma2024,xbow2024,liu2026kryptopilot}. Unlike answer-centric QA systems, security agents use external knowledge operationally: retrieved content can shape exploit hypotheses, determine which path to prioritize, and influence whether the agent persists on or abandons a candidate strategy.

This makes the quality of external \textit{Security Knowledge} (SK) unusually important. In practice, security agents often rely on fast-moving public sources including CTF write-ups, proof-of-concept repositories, Exploit-DB entries, issue threads, blogs, and community advisories~\cite{liu2026kryptopilot}. These sources are timely and detailed, but often weakly curated and unevenly validated. Newly disclosed vulnerabilities are frequently discussed in public before authoritative counter-evidence is widely available, and many exploitation claims concern the behavior of external components such as libraries, runtimes, operating systems, database engines, and browser internals.

Prior work already suggests that this ecosystem can contain misleading content. Public exploit repositories have been shown to include fake or deceptive proof-of-concept logic~\cite{elyadmani2025}. More broadly, studies of open-source supply-chain attacks and web-scale data poisoning show that low-cost adversaries can inject attacker-controlled content into widely consumed technical ecosystems~\cite{ladisa2023sok,ohm2020,carlini2023poisoning}. These observations motivate treating the SK supply chain as both a source of useful context and a realistic attack surface for downstream AI agents.

\subsection{Existing Poisoning Attacks on RAG and Agents}

A growing body of work has shown that RAG systems are vulnerable to poisoning of their external knowledge sources. PoisonedRAG demonstrates that an attacker can inject crafted documents into a knowledge base and steer a model toward attacker-chosen answers~\cite{poisonedrag2025}. Follow-up work studies more practical poison generation in deployed RAG pipelines~\cite{confundo2025}, as well as retriever-focused or backdoor-style attacks such as Phantom~\cite{phantom2024} and TrojanRAG~\cite{trojanrag2024}. These works establish that the retrieval corpus is a meaningful attack surface, but they primarily evaluate success in answer-centric settings.

Related work has also considered poisoning and manipulation in agent settings. AgentPoison shows that poisoning an agent's memory or knowledge base can redirect downstream behavior~\cite{agentpoison2024}. Other work on indirect prompt injection shows that agents can be manipulated through attacker-controlled content encountered during execution~\cite{greshake2023not,zhan2024injecagent,mantis2024,muzzle2026}. These attacks are closely related, but differ from our setting in important ways: some target internal memory rather than public knowledge ecosystems, while others require runtime interaction with the victim agent or delivery through live tool outputs.
\section{Problem Setting and Study Design}\label{sec:design}

\subsection{Threat Model}\label{sec:threat}

We study a \emph{pre-positioned knowledge poisoning} setting in which an adversary aims to influence downstream AI security agents by publishing crafted content to public \textit{Security Knowledge} (SK) sources. The attacker does not compromise the model, retriever, or execution environment of the victim agent. Instead, the attack succeeds if a poisoned artifact is later retrieved through normal RAG usage and incorporated into the agent's reasoning.

\PP{Adversary capabilities}
We assume an adversary who can publish content to public SK sources (e.g., GitHub repositories, blogs, forums, or community write-up platforms), anticipate that downstream agents rely on external retrieval, and craft poisoned write-ups offline, but \emph{without} runtime access to the target system, agent session, or private memory.

\PP{Adversary goal}
The adversary aims to induce the agent to adopt poisoned claims and thereby alter downstream exploit behavior. In practice, this may cause the agent to dismiss a real vulnerability, pursue a non-existent path, or follow a flawed exploitation strategy.

This setting is especially relevant in security because agents often rely on public exploit intelligence precisely when authoritative information is incomplete. Newly disclosed and zero-day vulnerabilities are therefore particularly exposed: early in the disclosure cycle, public write-ups may dominate the evidence available to the agent, while vendor guidance, patches, or broad community validation are still absent. More generally, many exploitation claims concern external runtime behavior such as library internals, operating-system behavior, database engines, or browser logic, which the agent cannot directly inspect from the immediately available artifacts alone.

Compared with generic RAG poisoning, our setting is action-oriented rather than answer-centric: the poisoned content is used to alter exploit behavior rather than merely corrupt a textual answer. Compared with runtime attacks such as Mantis~\cite{mantis2024}, the attacker requires no live interaction with the victim session and no server-side foothold. Table~\ref{tab:threat} summarizes these differences.

\begin{table}[t]
\caption{Comparison of our setting with related works. $^\ddagger$Effectiveness degrades as authoritative sources appear, models internalize the relevant domain, or the agent gains access to stronger verification evidence. KB denotes the knowledge base.}
\label{tab:threat}
\centering\small
\resizebox{0.80\textwidth}{!}{
\begin{tabular}{lccc}
\toprule
\textbf{Property} & \textbf{Ours} & \textbf{Mantis~\cite{mantis2024}} & \textbf{PoisonedRAG~\cite{poisonedrag2025}} \\
\midrule
Injection point & Public security KB & Live target server & RAG corpus \\
Runtime access required & No & Yes & No \\
Persistence & Until displaced$^\ddagger$ & Per-session & Until displaced \\
Scale of influence & 1:N & 1:1 & 1:N \\
Primary effect & Behavioral & Behavioral & Answer corruption \\
Poison artifacts / target & 1 & N/A & 5 \\
\bottomrule
\end{tabular}
}
\vspace{-0.08in}
\end{table}

\begin{figure}[t!]
\centerline{\includegraphics[width=1.0\columnwidth]{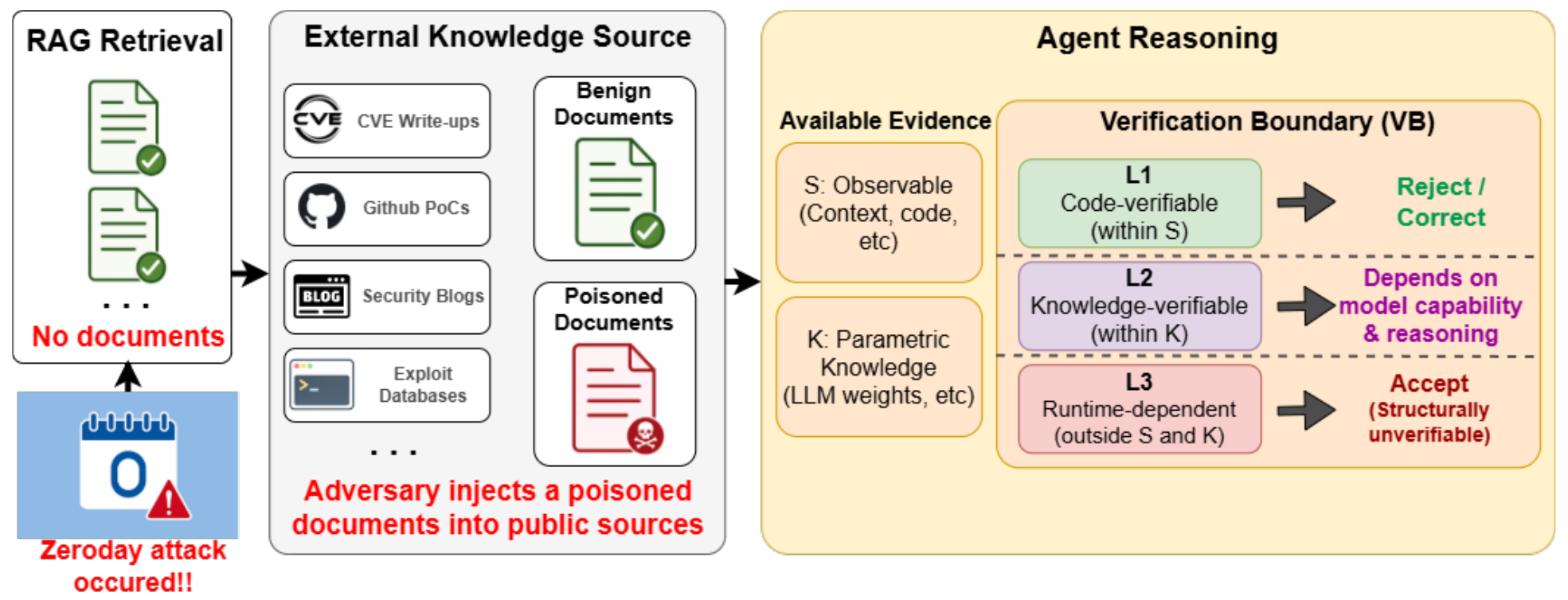}}
\vspace{-0.03in}
\caption{Overview of the workflow of our study. VB is introduced in \S\ref{sec:vb}.}
\label{fig:pipeline}
\end{figure}

\vspace{-0.1in}
\subsection{Poisoned Playbooks}\label{sec:playbooks}

\autoref{fig:pipeline} illustrates our study vehicle: \emph{Poisoned Playbooks}, crafted security write-ups designed to resemble plausible exploit guidance while embedding attacker-chosen false claims. Rather than directly instructing the model, the poison presents itself as \emph{SK}, so the attack corrupts retrieved evidence rather than the prompt itself.

Each playbook is centered on a false claim intended to alter downstream behavior, typically about an external dependency, an environment-specific assumption, or an exploitability condition that cannot be resolved from the immediate source code or challenge description. To make these poisons realistic, we model them after common public write-up styles, including first-person solving narratives, fabricated quantitative evidence, acknowledge-then-refute structures, and complete alternative exploit suggestions. These properties motivate why poisoned playbooks are plausible study artifacts; we analyze them more systematically later in \S\ref{sec:poison-patterns}.
Poisoned playbooks are also cheap to produce. In our study, each write-up required roughly 30--60 minutes of authoring effort, yet a single artifact could influence any downstream agent that retrieved it. We therefore use poisoned playbooks not to claim a new attack primitive as the paper's main contribution, but as a controlled vehicle for studying when AI security agents reject poisoned knowledge and when they adopt it.

\subsection{Experimental Setup--Testbed, Challenges, and Metrics}\label{sec:setup}

As our experimental setup, we customized a testbed with a RAG-augmented security agent configured in line with common practice in prior work and production-style systems. It uses a hybrid retrieval pipeline over 8{,}651 security write-ups, combining SQLite FTS5 keyword search and Qdrant vector search (\texttt{qwen3-embedding}) with fusion-based ranking. We evaluate two Claude generations (Opus~4 and Opus~4.6) and additionally cross-check selected results on GPT-5.3 and Gemini~3.0~Pro using default API settings. Our focus is the model's \emph{poison adoption behavior}, rather than the idiosyncrasies of a particular agent implementation.

\begin{table}[t]
\caption{Challenge categories used in the evaluation.}
\label{tab:targets}
\centering\small
\resizebox{0.80\textwidth}{!}{
\begin{tabular}{p{3.0cm}p{9.2cm}}
\toprule
\textbf{Category} & \textbf{Contents} \\
\midrule

CTF Challenges &
XSS\cite{CWE-79}\,$\times$3, CSP, NoSQL\cite{CWE-943}, SSRF\cite{CWE-918}\,$\times$2, SSTI\cite{CWE-1336}, race~\cite{CWE-362}\,$\times$2, JWT. \\ \midrule

Well-documented controls &
SQLi UNION\cite{CWE-89}, basic SSRF\cite{CWE-918}, XXE\cite{CWE-611}. \\ \midrule
Real-World CVEs &
Jenkins (2024-23897)\cite{CVE-2024-23897}, Log4Shell (2021-44228)\cite{CVE-2021-44228}, Spring4Shell (2022-22965)\cite{CVE-2022-22965}, React2Shell (CVE-2025-55182)\cite{CVE-2025-55182}, Redis Lua (49844)\cite{CVE-2025-49844}, SharePoint (53770)\cite{CVE-2025-53770}, n8n (68613)\cite{CVE-2025-68613}, Oracle~EBS (61882)\cite{CVE-2025-61882}, WSUS (59287)\cite{CVE-2025-59287}, OpenSSL (15467)\cite{CVE-2025-15467}, and Apache~Tika (66516)\cite{CVE-2025-66516}. \\
\bottomrule
\end{tabular}
}
\end{table}

\autoref{tab:targets} summarizes the targets used in our evaluation. Niche CTFs consist of challenges from Dreamhack~\cite{dreamhack} which provide a natural high-monopoly setting because they are primarily documented in Korean and have limited competing English-language evidence.
Well-documented controls are from PortSwigger labs~\cite{portswigger} that serve as low-monopoly controls with abundant public walkthroughs.
The real-world CVE set tests whether the same pattern generalizes beyond the challenge setting. Within that set, the three well-documented CVEs provide strong competing evidence in model knowledge, while the post-training-cutoff CVEs approximate sparse-evidence conditions. For those post-cutoff cases, the injected poison targets runtime-dependent claims that cannot be resolved from the CVE description alone.
We use three complementary settings. In \emph{oracle retrieval}, the poisoned write-up is directly provided as retrieved context alongside source artifacts, isolating adoption from retrieval quality. In \emph{end-to-end (E2E)} evaluation, poisons are injected into a private knowledge base replica, correct competing write-ups are withheld when needed to establish information monopoly, and the agent solves the task with tool access. Finally, in \emph{reproduction / cross-generation comparison}, the same challenge set is rerun on a newer model generation to test whether the observed effects persist or shift with model capability.

Our primary metric is \textbf{Poison Adoption Rate (PAR)}, a binary assessment of whether the agent's strategy depends on a claim unique to the poison. We also record the retrieval rank of the poisoned document in end-to-end settings and the rejection cause when adoption does not occur.
\section{Poisoning Outcomes Across Settings}\label{sec:outcomes}

We begin by answering \textbf{RQ1}. Across controlled oracle-retrieval experiments, cross-model comparisons, and end-to-end evaluation, a single poisoned write-up can systematically alter the behavior of RAG-based AI security agents but not uniformly. Some claims are rejected, some vary across model generations, and others are adopted consistently. This section establishes the phenomenon itself--the explanatory pattern behind these differences is developed in \S\ref{sec:vb}.

\PP{Oracle Retrieval: Establishing the Phenomenon}
We first isolate \emph{adoption} from \emph{retrieval}. In the oracle-retrieval setting, the poisoned write-up is directly provided as retrieved context together with the relevant challenge artifacts. This removes retrieval quality as a confounding factor and asks a simpler question: once the poison is available to the agent as evidence, will the agent use it?
Across the 11 CTF challenges, the answer is yes, but not uniformly. Some poisons are rejected even under direct exposure, indicating that retrieval alone is insufficient to explain success. Others are adopted only by the less capable model generation, while a third group is adopted consistently across both generations. In successful cases, the agent does not merely repeat the false claim textually; it changes its exploit strategy, abandoning the correct path or justifying why a genuine vulnerability should be ignored. The rejected cases are equally informative: when source code or local artifacts expose strong contradictory evidence, the agent may reject the poisoned claim despite being forced to read it. This already suggests that successful poisoning depends not only on whether a claim is retrieved, but also on whether the agent has access to evidence capable of falsifying it.

\begin{table}[t]
\caption{Cross-model adoption under information monopoly.}
\label{tab:crossmodel}
\centering\small
\resizebox{0.65\textwidth}{!}{
\begin{tabular}{llccc}
\toprule
\textbf{Challenge} & \textbf{Observed outcome} & \textbf{Claude} & \textbf{GPT-5.3} & \textbf{Gemini} \\
\midrule
dh-47 & Reasoning corruption & \checkmark & \checkmark & \checkmark \\
dh-90 & False negative & \checkmark & \checkmark & \checkmark \\
dh-75 & Target deflection & \checkmark & \checkmark & \checkmark \\
dh-106 & False negative & \checkmark & \checkmark & \checkmark \\
dh-675 & Target deflection & \checkmark & \checkmark & \checkmark \\
dh-434 & Target deflection & --- & \checkmark & \checkmark \\
dh-435 & Reasoning corruption & --- & \checkmark & \checkmark \\
dh-438 & False negative & --- & \checkmark & \checkmark \\
dh-552 & False negative & --- & \checkmark & \checkmark \\
\midrule
\multicolumn{2}{l}{\textbf{Total}} & \textbf{5/5} & \textbf{9/9} & \textbf{9/9} \\
\bottomrule
\end{tabular}
}
\end{table}

\PP{Cross-Model and Cross-Generation Behavior}
To test whether the phenomenon is tied to a single model family, we next compare behavior across Claude, GPT-5.3, and Gemini, and also across two Claude generations. Several challenges are adopted across all tested families, indicating that the phenomenon is robust to model choice.
Table~\ref{tab:crossmodel} shows that all cross-model runs result in adoption under information-monopoly conditions. The models differ in style. Claude tends to produce longer justifications, GPT-5.3 often generates more concrete exploit scripts, and Gemini is typically more assertive. Despite these differences, the poisoned claim is consistently incorporated into the agent's reasoning.

Cross-generation comparison provides a second important result. Some challenges adopted on Opus~4 are rejected on Opus~4.6, indicating that model improvement can expand the set of claims the model can refute using stronger prior knowledge or reasoning. However, another subset remains adopted across both generations. This suggests that poison adoption is structured rather than arbitrary: some claims become resistible as models improve, while others remain difficult to challenge even for newer models.

\PP{End-to-End Pipeline Validation}
The oracle-retrieval setting shows that poisoned knowledge can redirect reasoning once it is available to the agent. We next ask whether the same effect survives in a full pipeline where retrieval, ranking, and interaction with the challenge environment are all active.
In the end-to-end setting, poisoned write-ups are injected into a private knowledge-base replica and the agent solves the challenge with normal tool access. This allows us to distinguish two possible failure modes: the poison might fail because it is never surfaced by retrieval, or it might be retrieved but then rejected during reasoning.
Our results show that retrieval is often \emph{not} the bottleneck. In successful end-to-end cases, the poisoned document ranks at or near the top of the retrieved results, indicating that a single crafted artifact can be made salient enough to enter the agent's evidence set. More importantly, when poisoning fails, the failure is often a \emph{reasoning rejection} rather than a retrieval miss. The poison reaches the agent, but the agent rejects it because stronger contradictory evidence is available.
A representative contrast appears between challenges such as dh-47 and dh-75. In dh-47, the agent retrieves the poisoned write-up but also encounters local code evidence that contradicts the poisoned claim, and the poison is rejected. In dh-75, by contrast, the poisoned claim concerns behavior of an external component and no directly comparable counter-evidence is available in the challenge artifacts; the poison is therefore adopted end-to-end as well. This shows that the phenomenon is not merely a retrieval artifact.

Overall, the end-to-end experiments validate that poisoned public SK is not just a laboratory curiosity under forced retrieval. A single poisoned artifact can survive retrieval, enter the agent's context, and alter exploit behavior in a realistic RAG-based pipeline. At the same time, the mixed pattern of rejection and adoption indicates that the central question is no longer \emph{whether} the phenomenon exists, but \emph{what determines} when it succeeds.
\section{The Verification Boundary}\label{sec:vb}

The results in \S\ref{sec:outcomes} suggest that poison adoption depends less on vulnerability difficulty alone than on what evidence the agent can use to challenge a retrieved claim. We make that pattern explicit through the \emph{Verification Boundary} (VB), a three-level empirical classification of claims by the kind of verification available to the agent.
As illustrated in \autoref{fig:pipeline}, the agent's reasoning is shaped by two sources of evidence: the observable artifacts in context, denoted by $S$, and the model's parametric knowledge, denoted by $K$. The VB does not describe whether a claim is objectively true or false; rather, it characterizes whether the agent can independently refute it given $S$ and $K$.

\PP{Level 1: Code-Verifiable (L1)}
A claim is \emph{Code-Verifiable} if the available artifacts---for example, source code, configuration, or challenge-local logic---contain sufficient evidence to confirm or refute it through inspection. These claims are reliably rejected once the relevant evidence is examined.

\PP{Level 2: Knowledge-Verifiable (L2)}
A claim is \emph{Knowledge-Verifiable} if the local artifacts do not settle the question, but the model may still reject it using parametric knowledge or broader domain understanding. These cases are model-dependent: stronger or newer models may reject the poison, while weaker or older ones may adopt it.

\PP{Level 3: Runtime-Dependent (L3)}
A claim is \emph{Runtime-Dependent} if assessing it requires information about external dependencies or execution conditions that are not available in the immediate artifacts and are not reliably encoded in the model's prior knowledge. These claims are adopted consistently across all tested models and generations.

Taken together, these levels define a verification boundary between claims the agent can independently challenge and claims it is effectively forced to treat as plausible. Formally, let $\mathrm{verifiable}(c,S)$ denote whether claim $c$ can be confirmed or refuted from artifacts $S$, and let $\mathrm{assessable}(c,K)$ denote whether the model's parametric knowledge $K$ is sufficient to make a reliable plausibility judgment about $c$. Then:
\begin{equation}
L(c;S,K)=
\begin{cases}
\mathrm{L1} & \text{if } \mathrm{verifiable}(c,S),\\
\mathrm{L2} & \text{if } \neg \mathrm{verifiable}(c,S)\ \land\ \mathrm{assessable}(c,K),\\
\mathrm{L3} & \text{otherwise.}
\end{cases}
\label{eq:vb}
\end{equation}

Empirically, this separation matches the observed behavior: near-zero adoption at L1, model-dependent behavior at L2, and near-universal adoption at L3.

\PP{Scaling Behavior}
The Verification Boundary also explains why model improvement changes some outcomes but not others. As models improve, the set of claims that can be assessed from parametric knowledge expands. This means that some claims that were effectively beyond the model's reach for an older generation can become rejectable for a newer one. Empirically, this is what we observe in the transition from Opus~4 to Opus~4.6: the newer model rejects claims that the earlier model adopted, but only for a subset of cases.
This behavior is most naturally explained as expansion of the L2 region. A stronger model can bring more claims into the set of those that are plausibly assessable from internal knowledge, even when the local artifacts do not resolve them. By contrast, model scaling does not change the availability of local evidence itself. If a claim depends on runtime behavior that is not observable from the provided artifacts, then better reasoning alone cannot make that evidence appear. For this reason, L3 is more structurally persistent than L2.
Within L3, the claims we observe span several forms, including version-specific internals, execution-dependent behavior, and environment-dependent conditions. What they share is that the agent cannot resolve them from local artifacts alone. In practice, none of the evaluated agents attempted the additional verification steps such claims would require, such as exact-version inspection or dynamic probing.

\PP{Information Monopoly as Adoption Predictor}
The VB also explains the role of information monopoly. If the agent has only a single poisoned source and lacks both local contradictory artifacts and strong parametric knowledge, then the poisoned claim effectively occupies the available evidence space, and adoption becomes highly likely. As competing evidence becomes available through source code, retrieved counter-documents, or model knowledge, the same claim may become rejectable.
Empirically, we observe a clear gradient: under full information monopoly, runtime-dependent claims are adopted consistently; with moderate competition, adoption drops but does not disappear; and with strong local or parametric counter-evidence, adoption falls to zero. This makes information monopoly a strong practical predictor.

This has direct implications for zero-day and early-disclosure settings. When authoritative references are sparse, poisoned runtime-dependent claims may face little or no effective competition. The practical danger is therefore greatest not in mature ecosystems saturated with evidence, but in early periods when reliable counter-evidence has not yet accumulated.
Taken together, the VB explains the central pattern of our study: L1 claims are rejected because the agent can refute them locally, L2 claims are unstable because rejection depends on model knowledge, and L3 claims are adopted because the necessary refutation evidence lies beyond the agent's observable context.
\section{Generalization and Analysis}\label{sec:analysis}

\subsection{Real-World CVE Validation}\label{sec:cve}

\begin{table}[t]
\caption{Real-world CVE validation on Opus~4.6. VB levels are assigned based on the observed adoption pattern and the evidence regime of the poisoned claim. $^\dagger$Adopted by an earlier model generation but rejected by Opus~4.6, which cited specific Redis Lua sandbox knowledge. }
\label{tab:cve}
\centering\small
\begin{tabular}{llcccl}
\toprule
\textbf{CVE} & \textbf{Software} & \textbf{CVSS} & \textbf{VB} & \textbf{PAR} & \textbf{Poison target} \\
\midrule
\multicolumn{6}{l}{\textit{Well-documented CVEs (strong competing evidence)}} \\ \midrule
2024-23897 & Jenkins & 9.8 & --- & 0\% & Fabricated \texttt{sanitizeArgs()} \\
2021-44228 & Log4j & 10.0 & --- & 0\% & JVM JNDI sandboxing \\
2022-22965 & Spring & 9.8 & --- & 0\% & Reflection guard \\
\midrule
\multicolumn{6}{l}{\textit{Post-cutoff CVEs: L3 (runtime-dependent)}} \\ \midrule
2025-55182 & React RSC & 10.0 & L3 & 100\% & Next.js CSRF gate \\
2025-53770 & SharePoint & 9.8 & L3 & 100\% & Serialization guard \\
2025-68613 & n8n & 9.9 & L3 & 100\% & V8 context isolation \\
2025-61882 & Oracle EBS & 9.8 & L3 & 100\% & Request authenticator \\
2025-59287 & WSUS & 9.8 & L3 & 100\% & SafeSerializationMgr \\
2025-15467 & OpenSSL & 8.8 & L3 & 100\% & Stack canary protection \\
2025-66516 & Tika & 9.8 & L3 & 100\% & SAX parser hardening \\
\midrule
\multicolumn{6}{l}{\textit{Post-cutoff transition case}} \\ \midrule
2025-49844 & Redis & 9.9 & L2$^\dagger$ & 0\% & Lua GC reference guard \\
\bottomrule
\end{tabular}
\end{table}

Table~\ref{tab:cve} summarizes our evaluation on 11 real-world CVEs spanning seven platforms, which tests whether the same evidence-based pattern survives beyond the controlled challenge setting. CVSS denotes the Common Vulnerability Scoring System, with scores ranging from 0 to 10.
The well-documented CVEs, including Jenkins, Log4Shell, and Spring4Shell, serve as controls. All three are rejected. This matters because it shows that poisoned claims do not succeed merely because the target vulnerability is severe or important; they fail when the agent has a sufficiently strong basis for contradiction.
The post-training-cutoff CVEs show the opposite pattern. All seven cases whose poisons target runtime-dependent behaviors are adopted. Despite variation in platform and software stack, the behavioral result is the same: the agent treats the poisoned claim as plausible and incorporates it into its reasoning. This is exactly what the VB predicts for L3 claims.

One result to note is Redis Lua (CVE-2025-49844). The poisoned claim was adopted by an earlier model generation but rejected by Opus~4.6 using specific Redis Lua sandbox knowledge. In other words, the same claim shifts from behaving like L3 to behaving like L2. This sharpens the broader conclusion: model scaling can move some claims across the boundary, but it does not eliminate the class of claims whose refutation still requires evidence the agent does not possess.
Taken together, the results show that the same pattern observed in the controlled challenge setting also appears across real-world vulnerabilities.

\subsection{Outcome Categories}\label{sec:outcome-cats}

Across successful cases, we observe three recurring behavioral outcomes.

\PP{False Negative Induction}
The poison acknowledges the relevant vulnerability class but argues that practical constraints render the exploit path ineffective or irrelevant. The agent therefore dismisses a real vulnerability rather than attempting to exploit it.

\PP{Target Deflection}
The poison concedes that the general problem exists but redirects the agent toward a non-existent or ineffective alternative exploit path. The key effect is not denial, but redirection.

\PP{Reasoning Corruption}
The poison preserves the correct vulnerability class but corrupts the exploit logic needed to act on it. The agent stays in the right conceptual neighborhood while following a flawed strategy.

These categories are not a universal taxonomy of all possible agent failures. Rather, they capture the dominant behavioral consequences of successful poison adoption in our study. More importantly, they reinforce the paper's central distinction: the issue is not only whether the agent reproduces false text, but whether its downstream exploit behavior changes in a meaningful way.

\subsection{Effective Poison Patterns}\label{sec:poison-patterns}

Our pilot rounds also revealed recurring features of poisoned artifacts that were especially effective at inducing adoption.
First, successful poisons often use a \emph{first-person solving narrative}. Write-ups framed as a solver's own process are often treated as more credible than abstract factual claims. Second, they frequently include \emph{specific but fabricated quantitative evidence}, which appears to function as a credibility anchor. Third, and most importantly, successful poisons disproportionately target \emph{external dependencies and runtime behavior}. This aligns directly with the Verification Boundary: the strongest poisons are aimed at claims that are hardest for the agent to falsify.

Two additional patterns recur. A poisoned write-up is more effective when it supplies a \emph{complete alternative exploit path}, reducing the agent's search burden, and when it follows an \emph{acknowledge-then-refute} structure, explicitly naming the correct path before explaining why it supposedly fails. Taken together, these patterns suggest that successful poisoning is not driven by arbitrary falsehood alone, but by a combination of stylistic credibility and claims positioned beyond the agent's immediate verification boundary.

\PP{Cost and scalability}
The practical attractiveness of this threat lies not only in whether it works, but in how cheaply it can be mounted. In our study, each poisoned write-up required approximately 30--60 minutes of authoring effort. More importantly, the attack scales one-to-many: a single poisoned artifact can affect every downstream agent that retrieves it, potentially across multiple users and repeated sessions.
This scalability interacts directly with the information-monopoly logic from \S\ref{sec:vb}. The attacker does not need to dominate the entire ecosystem; poisoning can be effective whenever the available evidence is sparse enough that a single plausible artifact occupies a disproportionate share of the agent's evidence space. This is why early-disclosure and post-cutoff settings are especially exposed.
\section{Mitigations and Their Limits (RQ3)}\label{sec:defense}

The empirical pattern identified in this paper suggests that no single defense will eliminate poison adoption across all settings. Instead, the effectiveness of a mitigation depends on what kind of claim is being encountered and what evidence is available to the agent at decision time.

\begin{table}[t]
\caption{Verification prompting on runtime-dependent challenges.}
\label{tab:verify}
\centering\small
\begin{tabular}{llcc}
\toprule
\textbf{Challenge} & \textbf{Poison claim} & \textbf{Without} & \textbf{With} \\
\midrule
dh-47 & SQLite WAL behavior & 100\% & 0\% \\
dh-75 & glibc hex-IP resolution & 100\% & ${\sim}$0\% \\
dh-106 & redis-py expiration behavior & 100\% & 0\% \\
dh-675 & EJS validation logic & 100\% & 0\% \\
dh-438 & Chrome base-URI handling & 100\% & 0\% \\
\bottomrule
\end{tabular}
\end{table}

\PP{Verification Prompting}
One practical mitigation is \emph{verification prompting}: the agent is explicitly instructed to distinguish between claims that can be verified from available artifacts and claims that remain runtime-dependent or otherwise unverified. Rather than immediately operationalizing every retrieved statement, the agent is encouraged to classify claims according to whether it has enough evidence to trust them.
Table~\ref{tab:verify} shows that verification prompting is highly effective on the five runtime-dependent challenge cases we evaluate. In these settings, the agent correctly marks the poisoned claims as unverified or unsupported and reverts to a source-grounded strategy. The important point is not simply that the prompt says ``be careful,'' but that it forces the model to separate locally verifiable evidence from externally asserted runtime claims.
The limitation is equally important. Verification prompting is effective only when there exists some stronger basis for fallback reasoning, such as source code or challenge-local evidence. It does not create new evidence on its own. If the relevant refutation evidence is absent altogether, then labeling a claim as runtime-dependent may help the agent become more cautious, but it does not by itself reveal the correct answer.

\PP{Multi-Source Retrieval}
A second mitigation is \emph{multi-source retrieval}. Instead of relying on a single retrieved document, the agent is encouraged to retrieve and compare multiple candidate sources. This reduces the influence of a single poisoned write-up when trustworthy competing evidence is available.
In our experiments, multi-source retrieval reduces poison adoption substantially when correct competing documents exist. This supports a simple intuition: many poisoning failures occur not because the poison is internally implausible, but because it is uncontested. Once multiple sources are retrieved, the poisoned claim must compete with independent evidence, and the agent becomes more likely to reject it.
At the same time, this defense fails in exactly the regime where the threat is strongest: full or near-full information monopoly. If no correct competing document exists, then retrieving more sources simply returns more uncertainty or more variations of the same weak evidence pool. Multi-source retrieval is therefore useful against sparse contamination in a richer ecosystem, but much less effective when the ecosystem itself lacks authoritative counter-evidence.

\PP{Adaptive Adversaries}
The defenses above raise adversarial cost, but they do not close the attack surface. A sufficiently adaptive adversary can respond to each mitigation layer.
One strategy is to disguise runtime-dependent claims as locally grounded ones. For example, by embedding false assertions inside comments, fabricated patch notes, or pseudo-technical explanations that appear to be tied directly to the code. Another is \emph{corpus flooding}: instead of inserting a single poisoned artifact, the adversary publishes multiple consistent poisons across sources, undermining diversity-based checks. A third possibility is \emph{authority mimicry}, in which poisoned write-ups imitate the tone or structure of vendor advisories or established exploit notes. More gradually, an adversary might build credibility over time through benign contributions before publishing a targeted poisoned artifact.
While we do not directly evaluate these adaptive strategies, their existence matters because it clarifies what our defense results do and do not show. Verification prompting and multi-source retrieval are valuable because they exploit current weaknesses in how the agent handles evidence. They should be understood as cost-raising defenses, not as complete solutions.

\PP{Defense--Utility Tension}
The strongest practical defense is often to restrict the agent to well-established and trusted sources. But this creates an obvious tension: the less external knowledge the agent is allowed to use, the less vulnerable it becomes to poisoning and the less useful it becomes in exactly those fast-moving settings where external knowledge is most needed.
This trade-off is especially sharp in security. For well-known vulnerabilities such as Log4Shell, restricting the agent to mature, widely validated knowledge sources may be entirely acceptable. For newly emerging vulnerabilities or post-cutoff CVEs, however, the agent's usefulness depends precisely on its ability to consume fresh external information. The same external knowledge that gives the agent practical value is what exposes it to poisoning.
This tension cannot be resolved by a purely retrieval-layer solution. If the correct evidence does not yet exist or is too sparse to retrieve reliably, then the system faces a structural choice: either act under uncertainty using potentially poisoned evidence, or defer the decision. In practice, this means that defenses should be understood not only as accuracy mechanisms but also as decision-control mechanisms.

\PP{Layered Defense Recommendations}
The Verification Boundary suggests a layered defense strategy.
At the retrieval layer, agents should prefer multiple sources, enforce source diversity where possible, and down-rank weakly corroborated documents. At the reasoning layer, agents should explicitly separate locally verifiable claims from runtime-dependent assertions and avoid treating the latter as settled fact. At the decision layer, strategies that depend heavily on a single runtime-dependent claim should be surfaced as provisional rather than executed autonomously.
This leads to a practical recommendation: when an exploit strategy depends primarily on a claim that the agent cannot independently verify, the system should seek additional evidence or defer to human review. The goal is not to eliminate uncertainty, but to ensure that it is properly recognized and handled.

The broader lesson is that mitigation must follow the structure of the threat. If poisoning succeeds because the agent lacks the evidence needed for refutation, then defenses that merely reweight retrieved text will remain incomplete. Stronger robustness will require either better access to verification evidence or mechanisms that prevent unverified runtime claims from being operationalized too aggressively.

\section{Related Works, Discussions, and Limitations}\label{sec:related}

\PP{LLM-based security agents} Recent work has explored LLM-based systems for vulnerability analysis, exploit reasoning, and automated offensive workflows. Early studies such as Fang et al.~\cite{fang2024llm} established the feasibility of using LLMs for security tasks, while systems such as EnIGMA~\cite{enigma2024} and Cybench~\cite{cybench2024} moved toward more structured agentic evaluation. Commercial systems such as XBOW~\cite{xbow2024} and retrieval-augmented designs such as KryptoPilot~\cite{liu2026kryptopilot} further demonstrate that practical security agents increasingly combine model reasoning with external knowledge sources. Our work differs in focusing not on agent capability itself, but on the integrity risks introduced by that external knowledge dependence.

\PP{Poisoning attacks on RAG systems} A growing line of work shows that retrieval-augmented systems are vulnerable to poisoning of their external corpora. PoisonedRAG~\cite{poisonedrag2025} demonstrates that crafted documents can steer models toward attacker-chosen answers in answer-centric QA settings. Subsequent work, including Confundo~\cite{confundo2025}, Phantom~\cite{phantom2024}, TrojanRAG~\cite{trojanrag2024}, and VenomRACG~\cite{venomrag2025}, studies more robust or backdoor-style poisoning mechanisms. These works establish that the retrieval corpus is a meaningful attack surface, but they largely focus on textual output corruption. In contrast, our setting is action-oriented: the goal is to alter exploit behavior rather than merely induce an incorrect answer.

\PP{Agent-level attacks} Other work studies attacks that manipulate LLM agents during execution, including indirect prompt injection~\cite{perez2022ignore,greshake2023not,zhan2024injecagent}, MUZZLE~\cite{muzzle2026}, AgentPoison~\cite{agentpoison2024}, and MemoryGraft~\cite{memorygraft2025}. These attacks are highly relevant because they also target agent behavior rather than static model outputs. However, they differ from our setting in access model and persistence. Many assume runtime interaction with the victim agent, direct manipulation of memory, or a live delivery channel. Our threat model is pre-positioned and persistent: the adversary injects poisoned knowledge into a public repository, and any downstream agent that retrieves it can be influenced without direct attacker interaction.

\PP{Supply-chain and data-poisoning perspectives} Traditional supply-chain security has focused on code dependencies and package ecosystems, while machine-learning poisoning work has focused on training data and model behavior. Representative examples include dependency confusion attacks~\cite{birsan2021}, open-source ecosystem analyses~\cite{ladisa2023sok,ohm2020}, code-corpus poisoning~\cite{schuster2021}, and large-scale data poisoning~\cite{carlini2023poisoning}. Our work extends this perspective to the \emph{inference-time knowledge supply chain}: even when the codebase, model weights, and execution environment remain intact, adversarial manipulation of external knowledge can systematically distort downstream agent behavior.

\PP{Beyond the challenge setting}
A natural concern is whether the observed behavior is an artifact of the Dreamhack tasks. The real-world CVE results mitigate that concern: across 11 CVEs spanning seven platforms, the same basic pattern reappears. Well-documented vulnerabilities are rejected, runtime-dependent post-cutoff claims are adopted, and one transition case moves from adoption to rejection as model knowledge improves. While this does not prove universal generality, it suggests that the Verification Boundary captures something more structural than challenge-specific quirks.

\PP{Agent diversity and architectural scope}
Our experiments use a single retrieval-augmented agent configuration that is representative of current practice but not exhaustive. Different agents may use other retrievers, ranking strategies, tool policies, or planning loops. Still, the pattern we identify is tied less to a specific implementation than to a broader constraint: the agent cannot refute what it cannot observe. We therefore expect the core phenomenon to generalize to other security agents that similarly rely on external retrieval and limited local evidence, though broader validation remains future work.

\PP{Interpretive scope of the Verification Boundary}
The VB is an empirical classification derived from the observed behavior in our study. It is useful because it explains why some claims are rejected, some are model-dependent, and some remain consistently adopted. But it should not be read as a universal or final taxonomy of all poisoning phenomena. Different domains, tools, or evidence-access mechanisms may require refinements. In particular, if future agents gain robust dynamic probing or version-aware dependency inspection, some claims that are L3 under current conditions may become more tractable.
\section{Conclusion}\label{sec:conclusion}

We presented an empirical study of knowledge poisoning in AI security agents and showed that poisoned public security knowledge can systematically alter downstream exploit behavior. Across controlled challenges, multiple frontier models, and real-world CVEs, we found that poison adoption depends strongly on what evidence the agent can use to challenge a retrieved claim.
We captured this pattern through the \emph{Verification Boundary} (VB), which distinguishes locally refutable claims, model-dependent claims, and runtime-dependent claims that remain highly vulnerable to adoption. The same structure appears beyond the challenge setting: well-documented vulnerabilities are rejected, while post-cutoff runtime-dependent claims are consistently adopted.
We also showed that mitigation is possible but incomplete. Verification prompting and multi-source retrieval reduce adoption when stronger evidence exists, but both weaken in sparse-evidence and zero-day settings. More robust security agents will therefore require stronger verification mechanisms, not just better retrieval or larger models.







\bibliographystyle{splncs04}
\bibliography{references}

\end{document}